# Structural and Vibrational Properties of $D_3Se$ from First Principles: Anharmonic Quantum and Isotope Effects

Wenjie Ma [1†], Yao Ma [2†], Mi Pang [2], Pugeng Hou [1*] and Francesco Belli [3*]

1 *College of Science, Northeast Electric Power University, Changchun Road 169, Jilin 132012, People's Republic of China*
2 *Department of Applied Physics, School of Sciences, Xi'an University of Technology, Xi'an 710048, People's Republic of China*
3 *Oden Institute for Computational Engineering and Sciences, University of Texas at Austin, Austin, TX, USA.*

[†] **Wenjie Ma** and **Yao Ma** contribute equally to this work.

[*] Corresponding author. Email:

pugeng_hou@neepu.edu.cn; francesco.belli@austin.utexas.edu

## Abstract

Hydrogen-rich superconductors have garnered considerable interest following the discovery of hot superconductivity in high pressure $H_3S$, reviving prospects for room temperature superconductors under high-pressures. Using $H_3Se$ as a reference system, we investigate the vibrational and superconducting properties of $D_3Se$ in the $Im\overline{3}m$ phase across 60–200 GPa by combining first-principles calculations with the stochastic self-consistent harmonic approximation to treat ionic quantum and anharmonic effects. These effects introduce significant renormalization to the phonon spectra and stabilize the lattice down to at least ~70 GPa, well below the harmonic prediction of >110 GPa. Ultimately, the phonon renormalizations alter the electron–phonon coupling, introducing a decrease in the superconducting critical temperature by about 3–16 K across the studied pressure range with respect to standard calculations. Including anharmonic phonons within the Migdal–Eliashberg theory yields $T_c \approx 154$ K at 75 GPa (with $\mu^* = 0.1$, $\lambda = 3.0$), highlighting $D_3Se$ as a promising high-$T_c$ superconductor at moderate pressures. Examining the role of anharmonicity in the isotope effect, we find that at 200 GPa it suppresses the isotope coefficient $\alpha$ to ~0.29—one third below the harmonic value (~0.44) which approaches the BCS limit of ~0.5. This dramatic reduction demonstrates that anharmonicity fundamentally governs the isotope effect on this system. The stark discrepancy between anharmonic and harmonic descriptions underscores the need for targeted experimental efforts to resolve the origin of the persistent theory-experiment discrepancy in compressed hydrides.

## Introduction

The pursuit of superconductivity with high critical temperature ($T_c$) has long been a pivotal frontier in condensed matter physics. In this context, hydrogen-rich materials have played a significant role. In 1968, Ashcroft first proposed that compressing hydrogen to its metallic state could allow achieving high $T_c$ [1]; he later refined this hypothesis by suggesting that chemical pre-compression could reduce the required pressure to achieve high $T_c$ [2], although recent studies indicate that such high-$T_c$ behavior is unlikely to be realized at ambient pressure [3]. In the last decade, remarkable experimental breakthroughs in high-pressure hydrogen-rich compounds have proven such idea true, with reported $T_c$ values exceeding 200 K in systems such as sulfur hydride [4], lanthanum hydride [5,6], yttrium hydride [7,8], and calcium hydride [9,10] under pressures. Notably, $LaH_{10}$ currently holds the record with a $T_c$ of approximately 250 K at 150 GPa [5,6]. Alongside these experimental advances, first-principles calculations based on density functional theory (DFT) have established themselves as an indispensable predictive tool, having successfully anticipated numerous superconducting hydrides prior to their experimental synthesis [11–15] and providing crucial guidance for further exploration.

Within this rapidly evolving landscape of hydrogen-rich superconductors [16–24], a critical consensus has emerged: accurately predicting structural stability, lattice dynamics, and $T_c$ requires going beyond the standard harmonic approximation to explicitly incorporate quantum nuclear fluctuations and anharmonic effects [25–36]. This is particularly crucial for systems containing light elements like hydrogen, where substantial zero-point motion and pronounced anharmonicity can dramatically renormalize phonon spectra, stabilize otherwise unstable structures, and significantly enhance or suppress electron-phonon coupling and $T_c$. Notable examples include $LaH_{10}$, where quantum effects stabilize a high-$T_c$ structure [25]; $ScH_6$, where anharmonic stretching of $H_2$-like units boosts $T_c$ by approximately 15% [26]; and the molecular *Cmca* phase of hydrogen, in which anharmonicity doubles $T_c$ to above 200 K [27]. Conversely, in hydrides such as those of aluminum, palladium and platinum, anharmonic hardening of hydrogen-dominated optical modes suppresses superconductivity [28-30]. These cases underscore that quantum-anharmonic effects are not mere corrections but are often decisive in governing the physical behavior of hydrogen-rich superconductors [36].

In this context, the isoelectronic selenide analogue of the well-established $H_3S$ system—$H_3Se$—offers a compelling yet underexplored platform. While $H_3Se$ has garnered theoretical interest, existing studies, reporting $T_c$ values ranging from ~110 K to ~163 K within the cubic $Im\bar{3}m$ phase [37–38], have predominantly relied on harmonic DFT calculations and omitted quantum-anharmonic treatments. Recent work demonstrates that incorporating these effects via the stochastic self-consistent harmonic approximation (SSCHA) significantly lowers the dynamic stability pressure of $H_3Se$ to ~61 GPa, substantially modifies its optical phonon modes, and yields a high $T_c$ of ~200 K around 64 GPa [39]. These findings not only underscore the critical importance of beyond-harmonic methodologies but also raise fundamental questions regarding isotopic substitution. For instance, anharmonic effects driven by quantum fluctuations can introduce deviations from the expected behavior for isotopic substitutions and, in certain cases such as PdH, can even produce an inverse effect where heavier isotopes increase $T_c$ [29, 40-42]. For this reason, it is critical to pose the question: how do quantum-anharmonic effects manifest in the deuterated $D_3Se$, and what is their combined impact on its structural, vibrational, and superconducting properties? Furthermore, how does anharmonicity influence the isotope coefficient α—a key parameter linking theory and experiment in such systems? Addressing these questions is essential not only for advancing the understanding of selenide-based superconductors but also for elucidating general trends in quantum-anharmonic lattice dynamics across hydrides and deuterides.

In this work, we present a comprehensive first-principles study of the structural, vibrational, and superconducting properties of $D_3Se$ in the $Im\bar{3}m$ phase over a broad pressure range (60 ~ 200 GPa), with particular focus on quantum-anharmonic and isotope effects. By employing the SSCHA to treat ionic quantum fluctuations and anharmonicity nonperturbatively, and by systematically comparing with recently reported $H_3Se$ results [39], we achieve two main objectives: (1) to determine how quantum-anharmonic effects modify the phase stability, lattice dynamics, and electron-phonon coupling in $D_3Se$; (2) to quantify the isotopic dependence of these properties and evaluate the resulting isotope coefficient. Our calculations reveal that quantum-anharmonic corrections significantly stabilize the $D_3Se$ lattice, lower its dynamic stability threshold relative to harmonic predictions, and induce substantial renormalization of the phonon spectrum and Eliashberg spectral function. These modifications lead to a suppression of $T_c$ compared to harmonic estimates, while still preserving high-$T_c$ behavior (~152 K) at 75 GPa when anharmonic phonons are considered within the Migdal–Eliashberg framework. Furthermore, we find that anharmonicity reduces the isotope coefficient $\alpha$ below the canonical BCS value, highlighting a nontrivial interplay between lattice anharmonicity and isotopic mass substitution. This study not only advances the understanding of high-temperature superconductivity in deuterated selenides but also underscores the necessity of incorporating quantum-anharmonic lattice dynamics for reliable predictions in hydrogen-rich superconductors, particularly in the context of isotopic studies.

**Methodology**

**A. First-Principles and Lattice Dynamics Framework**

Within the conventional ab initio framework, vibrational properties are typically obtained within the harmonic approximation via density functional perturbation theory (DFPT) [43]. In this approach, ions are treated as classical particles, and the Born-Oppenheimer (BO) potential energy surface $V(\boldsymbol{R})$ is expanded to second order around its minimum configuration $\boldsymbol{R}_0$. The harmonic phonon frequencies $\omega_{\nu}^{(h)}(\boldsymbol{q})$ are derived from the eigenvalues of the classical harmonic dynamical matrix:

$$D_{ab}^{(h)} = \frac{1}{\sqrt{M_a M_b}} \frac{\partial^2 V(\boldsymbol{R})}{\partial R_a \partial R_b}\bigg|_{\boldsymbol{R}_0}. \quad (1)$$

where the combined indices *a, b* run over all atomic Cartesian coordinates, $M_a$ is the atomic mass, and $\nu$ and $\boldsymbol{q}$ denote the phonon branch and wavevector, respectively.

However, for hydrogen/deuterium-rich compounds like $D_3Se$, significant quantum zero-point motions and pronounced anharmonicity of $V(\boldsymbol{R})$ render the harmonic approximation inadequate. It often incorrectly estimates the pressure required for dynamical stability and provides an inaccurate description of the phonon spectrum, which directly impacts derived superconducting properties. To accurately capture these essential quantum anharmonic effects, we employ the SSCHA.

**B. The Stochastic Self-Consistent Harmonic Approximation (SSCHA)**

The SSCHA is a variational, quantum-statistical method that incorporates ionic quantum fluctuations and anharmonicity without resorting to a perturbative expansion of $V(\boldsymbol{R})$ [29,30,44]. Its core principle involves minimizing the Gibbs-Bogoliubov free energy of the system with respect to a trial density matrix $\tilde{\rho}(\boldsymbol{\mathcal{R}}, \boldsymbol{\Phi})$ parametrized by centroid positions $\boldsymbol{\mathcal{R}}$ and auxiliary force constants $\boldsymbol{\Phi}$:

$$F[\tilde{\rho}(\boldsymbol{\mathcal{R}}, \boldsymbol{\Phi})] = \langle K + V(\boldsymbol{R})\rangle_{\tilde{\rho}(\boldsymbol{\mathcal{R}},\boldsymbol{\Phi})} - TS\,[\tilde{\rho}(\boldsymbol{\mathcal{R}}, \boldsymbol{\Phi})]. \tag{2}$$

Here, $K$ is the ionic kinetic energy operator, $T$ is the temperature, and $S\,[\tilde{\rho}(\boldsymbol{\mathcal{R}}, \boldsymbol{\Phi})]$ is the entropy. The centroids $\boldsymbol{\mathcal{R}}$ define the average ionic positions, while $\boldsymbol{\Phi}$ governs the width of the quantum-thermal fluctuations around them, representing an effective, temperature-dependent potential.

The SSCHA code minimizes $F[\tilde{\rho}(\boldsymbol{\mathcal{R}}, \boldsymbol{\Phi})]$ with respect to both $\boldsymbol{\mathcal{R}}$ and $\boldsymbol{\Phi}$, simultaneously optimizing internal coordinates and lattice parameters under a target external pressure. This procedure fully includes quantum and anharmonic effects at the specified conditions. The converged centroid positions $\boldsymbol{\mathcal{R}}_{\mathrm{eq}}$ represent the quantum-anharmonic equilibrium structure, which generally differs from the classical minimum $\boldsymbol{R}_0$. The corresponding auxiliary force constants $\boldsymbol{\Phi}_{\mathrm{eq}}$ characterize the effective vibrational behavior on the true anharmonic potential surface.

To obtain the phonon excitation spectrum corresponding to this anharmonic ground state, we compute the quantum-anharmonic dynamical matrix from the second derivative (Hessian) of the free energy at $\boldsymbol{\mathcal{R}}_{\mathrm{eq}}$:

$$D_{ab}^{(F)} = \frac{1}{\sqrt{M_a M_b}} \frac{\partial^2 F}{\partial \boldsymbol{\mathcal{R}}_a \partial \boldsymbol{\mathcal{R}}_b}\bigg|_{\boldsymbol{\mathcal{R}}_{\text{eq}}}, \tag{3}$$

The eigenvalues $\omega_\nu^{(F)}(\boldsymbol{q})$ of $\boldsymbol{D}^{(F)}$ yield the anharmonic phonon frequencies in the static limit. A negative eigenvalue (imaginary frequency) signifies dynamical instability within the quantum-anharmonic free energy landscape, a more rigorous criterion than that based on the classical matrix $\boldsymbol{D}^{(\text{h})}$. In practice, $\boldsymbol{D}^{(F)}$ is efficiently computed via finite differences using ensembles of stochastically generated ionic configurations sampled from the optimized $\tilde{\rho}(\boldsymbol{\mathcal{R}}, \boldsymbol{\Phi})$, $\boldsymbol{\Phi}_{\text{eq}}$.

Beyond optimizing internal ionic positions by minimizing the free-energy functional F[$\tilde{\rho}(\boldsymbol{\mathcal{R}}, \boldsymbol{\Phi})$, $\boldsymbol{\Phi}_{\text{eq}}$] with respect to the centroids, the SSCHA is also capable of relaxing lattice cell parameters while accounting for quantum nuclear effects and anharmonicity. This is accomplished by evaluating the stress tensor as the derivative of the SSCHA free energy with respect to the components of the strain tensor $\epsilon$:

$$P_{\alpha\beta} = -\frac{1}{\Omega_V} \frac{\partial F}{\partial_{\epsilon_{\alpha\beta}}} |_{\epsilon = 0}, \tag{4}$$

where $\Omega_V$ is the volume of the simulation supercell, and $\alpha$, $\beta$ denote Cartesian indices. This formulation enables the inclusion of additional pressure contributions arising from ionic quantum fluctuations, beyond those captured within the classical harmonic approximation—the latter being obtained when replacing the SSCHA free energy in Eq. (4) with the Born–Oppenheimer potential $V(\mathbf{R})$. Consequently, this approach provides a more complete description of the additional pressure contributions arising from ionic quantum fluctuations.

**C. Calculation of Superconducting Properties and the Isotope Effect**

Using the anharmonic phonon spectra $\omega_{\nu}^{(F)}(\boldsymbol{q})$ and eigenvectors from the SSCHA, we calculate the electron-phonon coupling (EPC) spectral function $\alpha^2F(\omega)$ and the total EPC constant $\lambda$:

$$\lambda = 2\int_0^{\infty} \frac{\alpha^2F(\omega)}{\omega} d\omega. \quad (5)$$

The superconducting critical temperature $T_c$ is determined by solving the Allen-Dynes modified McMillan equation[45],

$$T_c = \frac{f_1 f_2 \omega_{log}}{1.2} \exp\left[-\frac{1.04(1+\lambda)}{\lambda-\mu^*(1+0.62\lambda)}\right], \quad (6)$$

Here $\mu^*$ is the effective Coulomb potential, and the other parameters are calculated as follows:

$$\omega_{log} = \exp\left[\frac{2}{\lambda}\int_0^{\infty} \frac{d\omega}{\omega} \alpha^2F(\omega) \ln\omega\right], \quad (7)$$

$$f_1 = \sqrt[3]{\left[1+\left(\frac{\lambda}{\Lambda_1}\right)^{\frac{3}{2}}\right]}, \quad (8)$$

$$f_2 = 1 + \frac{(\frac{\bar{\omega}_2}{\omega_{log}}-1)\lambda^2}{\lambda^2+{\Lambda_2}^2}, \quad (9)$$

$\Lambda_1$, $\Lambda_2$ and $\bar{\omega}_2$ are given by

$$\Lambda_1 = 2.46(1+3.8\mu^*), \quad (10)$$

$$\Lambda_2 = 1.82(1+6.3\mu^*)\frac{\bar{\omega}_2}{\omega_{log}}, \quad (11)$$

$$\bar{\omega}_2 = \sqrt{\frac{2}{\lambda}\int_0^{\infty} \alpha^2F(\omega)\,\omega\, d\omega}, \quad (12)$$

We calculate the Eliashberg function as

$$\alpha^2F(\omega) = \frac{1}{2N(0)N_qN_k}\sum_{\boldsymbol{k}nm,\nu\boldsymbol{q},\bar{a}\bar{b}} \frac{\epsilon_\nu^{\bar{a}}(\boldsymbol{q})\epsilon_\nu^{\bar{b}}(\boldsymbol{q})^*}{\omega_\nu(\boldsymbol{q})\sqrt{M_{\bar{a}}M_{\bar{b}}}} \times d^{\bar{a}}_{\boldsymbol{k}n,\boldsymbol{k}+\boldsymbol{q}m} d^{\bar{b}^*}_{\boldsymbol{k}n,\boldsymbol{k}+\boldsymbol{q}m}\delta(\varepsilon_{\boldsymbol{k}n})\delta(\varepsilon_{\boldsymbol{k}+\boldsymbol{q}m})\delta(\omega-\omega_\nu(\boldsymbol{q})) .$$

(13)

In the equation above, $d^{\bar{a}}_{\boldsymbol{k}n,\boldsymbol{k}+\boldsymbol{q},m} = <\boldsymbol{k}n\left|\frac{\delta V_{KS}}{\delta R^{\bar{a}}(\boldsymbol{q})}\right|\boldsymbol{k}+\boldsymbol{q},m>$, where $|\boldsymbol{k}n>$ represents a Kohn-Sham state with energy $\varepsilon_{\boldsymbol{k}n}$ measured from the Fermi level, $V_{KS}$ denotes the Kohn-Sham potential, and $R^{\bar{a}}(\boldsymbol{q})$ stands for the Fourier-transformed displacement of atom $\bar{a}$; The combined atom and Cartesian indices with a bar ($\bar{a}$) only run for atoms within the unit cell. $N_k$ and $N_q$ are the number of electron and phonon momentum points utilized for BZ sampling; $N(0)$ represents the density of states at the Fermi level, while $\omega_\nu(\boldsymbol{q})$ and $\epsilon_\nu^{\bar{a}}(\boldsymbol{q})$ represent phonon frequencies and the polarization vectors, respectively. In this study, the Eliashberg function is computed both at the harmonic and anharmonic levels, by substituting into Eqs. (13) the harmonic phonon frequencies and polarization vectors obtained by diagonalizing $\boldsymbol{D}^{(h)}$ or their anharmonic equivalents from diagonalizing $\boldsymbol{D}^{(F)}$. It is important to note that the derivatives of the Kohn–Sham potential used in the electron–phonon matrix elements are evaluated at different positions in classical harmonic and quantum anharmonic calculations: in the former, they are computed at the positions $\boldsymbol{R}_0$ minimizing $V(\boldsymbol{R})$, while in the latter, they are determined at the positions $\boldsymbol{\mathcal{R}}_{\text{eq}}$ that minimize instead $\mathcal{F}[\tilde{\rho}(\boldsymbol{\mathcal{R}}, \boldsymbol{\Phi})]$. For comparison, $T_c$ is also determined from solving the isotropic Migdal-Eliashberg equations once $\alpha^2F(\omega)$ is obtained[46-47].

$$\Delta_n = \pi T_c \sum_m \frac{\lambda(n-m)-\mu^*}{\sqrt{{\omega_m}^2+\Delta_m^2}} \Delta_m, \tag{14}$$

where $\Delta_n$ is the gap function at the Matsubara frequency $\omega_n=(2n+1)\pi T_c$ , $\lambda(n-m)$ is the EPC kernel transformed from $\alpha^2F(\omega)$, and $\mu^*$ is the effective Coulomb pseudopotential[48]. We adopt a standard range of $\mu^*$=0.10−0.15 for our calculations.

The isotope coefficient $\alpha$ was calculated using the standard expression, where the mass ratio refers specifically to that of the substituted atoms (D/H), as the superconducting properties are predominantly governed by vibrations involving the hydrogen/deuterium sublattice. The isotope coefficient $\alpha$ is then computed from the calculated critical temperatures[49]:

$$\alpha = -\frac{\ln\{T_C(\mathrm{H_3Se})/T_C(\mathrm{D_3Se})\}}{\ln\{\mathrm{M(H)/M(D)}\}} \tag{15}$$

By comparing the $\alpha$ values obtained from purely harmonic calculations with those from full SSCHA anharmonic calculations, we can quantitatively assess the role of anharmonicity in enhancing or suppressing the isotope effect—a key fingerprint connecting theory and experiment in superconducting hydrides/deuterides. All necessary third-order interatomic force constants for the SSCHA are extracted from DFPT calculations performed on appropriately large supercells to ensure accuracy.

**Computational Details**

The calculations were conducted within the ab initio framework of the QUANTUM ESPRESSO package [50-51]. The exchange-correlation potential was treated using the Perdew–Burke–Ernzerhof parametrization [52] in conjunction with ultrasoft pseudopotentials[53]. Wavefunction and charge density cutoff energies were set to 80 Ry and 800 Ry, respectively. For the self-consistent electronic structure calculations, Brillouin zone integration was performed using a 24×24×24 ***k***-point grid with Methfessel–Paxton smearing (broadening width of 0.01 Ry). Harmonic phonon properties were obtained via density functional perturbation theory (DFPT) [43] on a 9×9×9 ***q***-point grid. The SSCHA calculations employed a 3×3×3 supercell containing 108 atoms. The resulting anharmonic dynamical matrices were then interpolated onto a finer 9×9×9 ***q***-point grid for analysis. Electron-phonon coupling was evaluated in both the harmonic and anharmonic regimes, using a Gaussian smearing of 0.024 Ry for the calculations. It should be noted that all computational parameters adopted here are identical to those utilized in the calculations for $H_3Se$.

## Results and Discussion

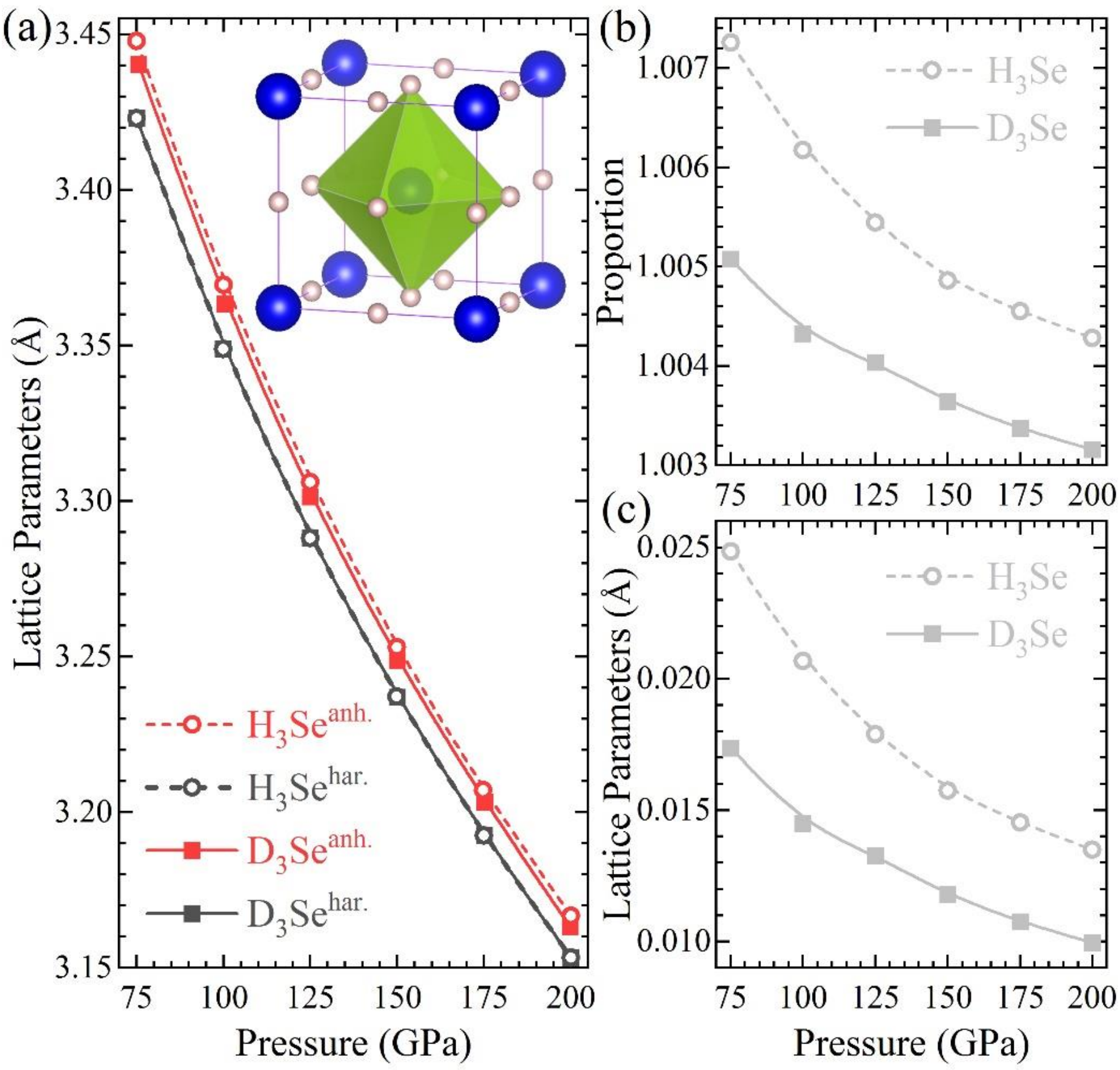

FIG. 1. (a) Lattice parameter as a function of pressure for the cubic $Im\overline{3}m$ phases of $H_3Se$ (dotted) and $D_3Se$ (solid). Red symbols and lines represent the anharmonic results obtained from the SSCHA free energy in the fixed−target−pressure mode, while black symbols and lines correspond to the harmonic (classical) approximation evaluated with the Born–Oppenheimer potential. The two harmonic curves coincide exactly, reflecting the mass independence of the classical limit. Panels (b) and (c) show the ratio and difference of the anharmonic to harmonic lattice parameters at each pressure for $H_3Se$ (dotted) and $D_3Se$ (solid), respectively, to clearly illustrate the lattice expansion induced by anharmonic effects.

$H_3Se$ has attracted recent research interest due to its stability and properties under high pressure. Previous study [38] indicate that this compound can be stabilized in the highly symmetric $Im\overline{3}m$ phase above 64 GPa. Moreover, in calculations using the Migdal-Eliashberg equations, its $T_c$ reaches 200 K. Among all covalent compounds, this represents one of the highest superconducting transition temperatures achieved at moderate pressures.

In the present work, we focus on this phase and systematically investigate the influence of quantum ionic fluctuations on the structural and vibrational properties by substituting H with its isotope deuterium (D). As shown in Fig. 1 inset, the phase has a body-centered symmetric structure: each Se atom is octahedrally coordinated by six D atoms, and each D atom has a similar environment. By performing both harmonic and anharmonic calculations, we examine the structural deformation under applied pressure. The pressure dependence of the lattice parameter is shown in Fig. 1(a), where the data for $H_3Se$ are also included for the convenience of isotope analysis. The two harmonic curves for $H_3Se$ and $D_3Se$ coincide exactly, reflecting the mass independence of the classical limit: within the harmonic approximation (without zero-point energy corrections), the atomic equilibrium positions $\boldsymbol{R}_0$ are determined solely by the Born-Oppenheimer potential, which is independent of the isotopic mass. For $D_3Se$, the introduction of quantum and anharmonic effects does not alter the overall crystal structure, but merely induces a minor correction to the lattice parameter. Specifically, at a given interionic stress, the harmonic lattice parameter is only about 0.010–0.018 Å smaller than that obtained from the SSCHA, indicating a very slight lattice contraction. Moreover, the slope of the equation of state (EOS) curve remains almost unaffected. When this tiny lattice constant correction is converted into a pressure correction at a fixed lattice constant, it yields a pressure shift of approximately 6 GPa.

This modest shift is consistent with the expected quantum pressure effects and further confirms that the anharmonicity introduces only a minor quantitative adjustment without altering the structural properties. Such pressure correction is a commonly observed feature in superhydrides, such as $H_3Se$[38] and $AlH_3$[27]. The pressure correction for $H_3Se$ is about 8 GPa, which is slightly larger than the 6 GPa for $D_3Se$. This difference can be attributed to the fact that quantum anharmonic effects are weaker for the heavier D compared to H. A clearer illustration is provided in Fig. 1(b) and (c), which present, respectively, the pressure-dependent ratios and differences between the anharmonic lattice constants (obtained from SSCHA calculations) and their harmonic counterparts for both $H_3Se$ and $D_3Se$.

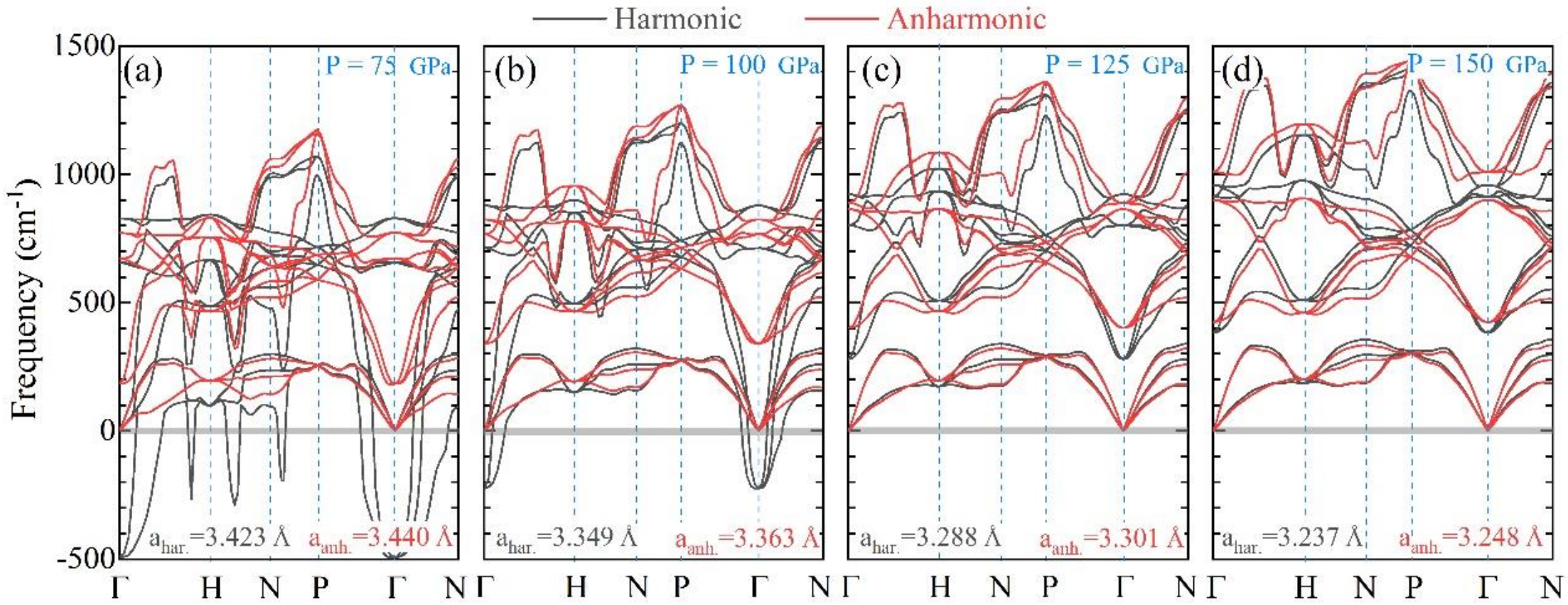

FIG. 2. Phonon spectra of the cubic $Im\overline{3}m$ $D_3Se$ at selected pressures, comparing harmonic (black) and anharmonic (red) results: (a) 75 GPa, (b) 100 GPa, (c) 125 GPa, and (d) 150 GPa. The anharmonic spectra, derived from the static limit of the SSCHA dynamical theory via the Hessian of the free energy $\boldsymbol{D}^{(F)}$ in Eq (3), while the harmonic results are obtained from the classical harmonic dynamical matrix $\boldsymbol{D}^{(h)}$ in Eq. (1). A solid grey line separates the regions of positive frequencies from those of negative (imaginary) frequencies.

In addition to the structural corrections, quantum anharmonic effects also significantly influence the phonon spectrum and dynamical stability of $D_3Se$. Figure 2 compares the phonon spectra of $D_3Se$ in the $Im\overline{3}m$ phase calculated within the harmonic approximation (black dotted lines) and the SSCHA (red solid lines) at pressures of 75, 100, 125, and 150 GPa (panels (a)–(d), respectively).

Consistent with trends observed in other hydrides, harmonic calculations at lower pressures (below 100 GPa, Figs. 2(a) and 2(b)) predict structural instability, evidenced by imaginary frequencies, particularly at the Γ point—a finding in general agreement with prior work [38]. In stark contrast, the anharmonic phonon spectrum remains entirely real above 75 GPa, demonstrating the critical stabilizing role of quantum ionic fluctuations in this deuterated system.

The phonon behavior of $D_3Se$ closely resembles that of $H_3Se$[38]. Above 100 GPa, quantum ionic and anharmonic fluctuations lead to a separation between the acoustic and optical branches. The acoustic modes, which are predominantly governed by selenium atom vibrations, show negligible variations. In contrast, the optical phonons exhibit notable renormalization: the six middle-frequency bond-bending modes are generally softened (except at the Γ point), while the three highest-frequency bond-stretching modes are hardened. This differentiated response of bending and stretching modes to anharmonicity is consistent with previous theoretical analyses of hydrogen-rich compounds [32,39], and similar band-mixing features between these optical branches are observed at lower pressures for both isotopic compounds.

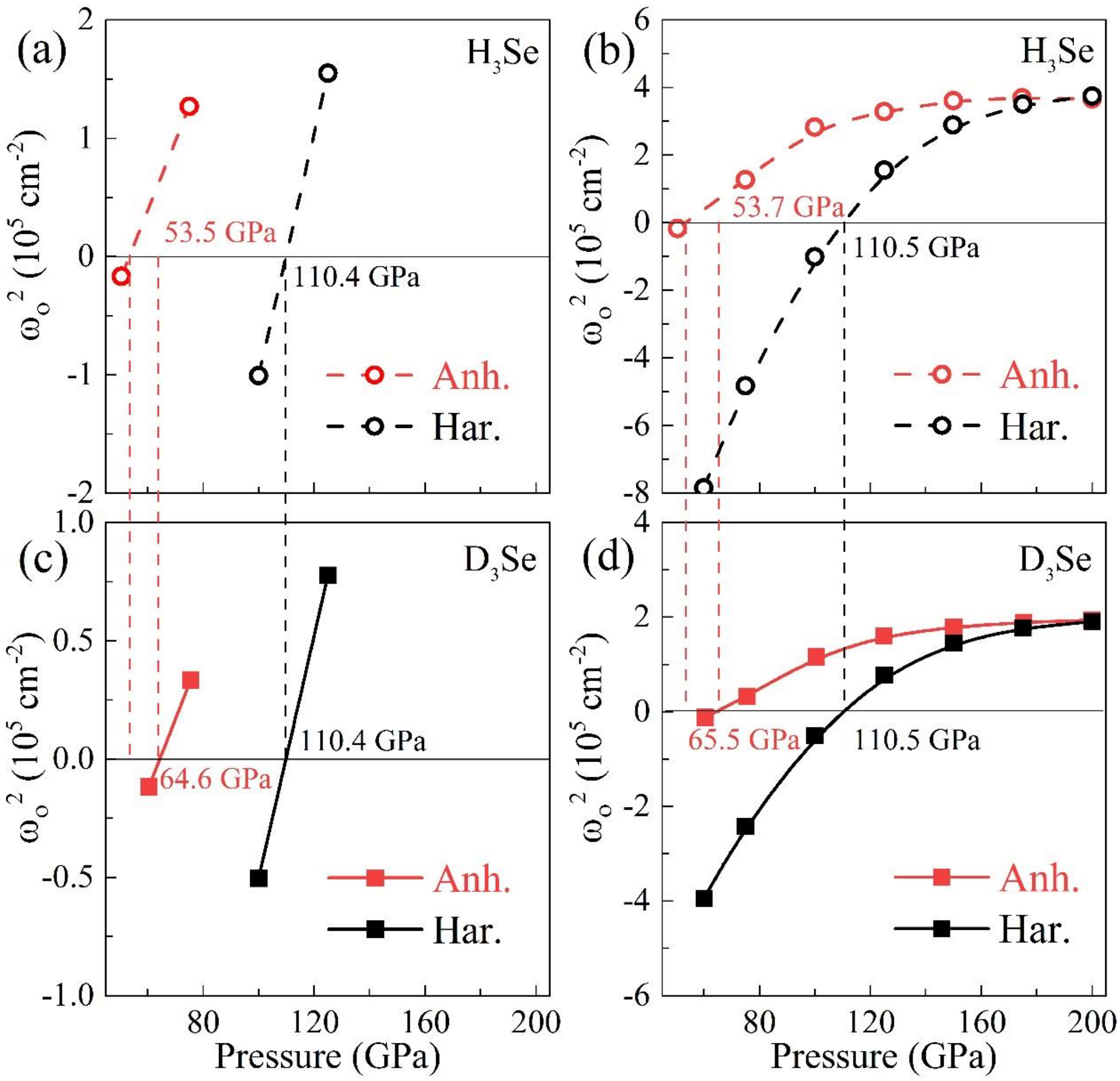

FIG. 3. Squared frequency $\omega_0^2$ of the $T_{1u}$ mode at Γ for $H_3Se$ and $D_3Se$ as a function of pressure. The color and line-style conventions are the same as in Fig. 1(a): red/black for anharmonic/harmonic results, and solid/dashed for $D_3Se$/$H_3Se$, respectively. Panels (a) and (c) focus on the two points immediately bracketing the zero crossing (with linear fits) to extract the transition pressure. Panels (b) and (d) present the full pressure dependence of $\omega_0^2$, illustrating the overall trend and validating the reliability of the linear extrapolation. The anharmonic results are calculated by diagonalizing the free energy dynamical matrix $\boldsymbol{D}^{(F)}(\boldsymbol{q})$ [Eq. (3)] and the harmonic results by diagonalizing the harmonic dynamical matrix $\boldsymbol{D}^{(h)}(\boldsymbol{q})$ [Eq. (1)].

To examine the dynamical stability of the $Im\overline{3}m$ phase in $D_3Se$ relative to $H_3Se$, we follow the methodology established for the $H(D)_3S$ [50]. In that work, the critical pressure for the $Im\overline{3}m \rightarrow R3m$ transition was found to be identical for both isotopologues within the classical harmonic approximation (173 GPa), because the condition for the harmonic dynamical matrix $\boldsymbol{D}^{(h)}$ in Eq. (1) to yield imaginary frequencies depends solely on the second derivatives of the Born-Oppenheimer $V(\boldsymbol{R})$ potential at $\boldsymbol{R}_0$, which are independent of the isotopic mass. However, when quantum anharmonic effects are included via the SSCHA, both the free energy $F$ and the renormalized atomic positions $\boldsymbol{R}_{\mathrm{eq}}$ in the dynamical matrix $\boldsymbol{D}^{(F)}$ in Eq.(3) become mass-dependent. As a result, a pronounced isotope effect emerged: the critical pressure dropped to 91 GPa for $H_3S$ and 107 GPa for $D_3S$.

Applying the same analysis to our calculated data for the Se-based analogue (Fig. 3), we define $\omega_0$ as the frequency of the $T_{1u}$ mode at the Γ point for $H_3Se$/$D_3Se$ in the $Im\bar{3}m$ phase and observe an analogous behavior. Within the harmonic approximation, fitting $\omega_0^2$ yields essentially identical critical pressures for $H_3Se$ and $D_3Se$, with a difference of merely ~0.1 GPa. This near-equivalence reflects the fact that, similar to the case of $H_3S$/$D_3S$ mentioned above, at the harmonic level the critical pressure is independent of the isotopic mass for $H_3Se$/$D_3Se$. In stark contrast, once anharmonicity and quantum fluctuations are fully considered, the critical pressures differs substantially. As clearly shown in Fig.3(c) and (d), for $D_3Se$, the critical pressure drops significantly from approximately 110 GPa in the harmonic approximation to about 65 GPa when quantum anharmonic SSCHA corrections are included. This reduction highlights the significant impact of quantum anharmonic effects on the dynamical stability, which is consistent with our findings for $H_3Se$ [39].

Furthermore, a comparison between $H_3Se$ and $D_3Se$ reveals an isotope effect on the critical pressure that cannot be captured by the harmonic approximation. From the anharmonic $\omega_0^2$ data, we obtain a difference of 11~12 GPa between $H_3Se$ and $D_3Se$. Although these values are smaller than the 16 GPa isotope shift reported for the sulfide system [54], they still indicate the same underlying physics operating in the selenide—namely, the mass-dependent renormalization of the potential energy surface by zero-point motion and anharmonic fluctuations. This underscores that an accurate description of the phase diagram of compressed hydrides (and deuterides) must go beyond the harmonic approximation to capture the essential roles of anharmonicity and isotopic mass. Notably, the consistently higher stabilization pressure of $D_3Se$ suggests that deuteration shifts the system toward the harmonic limit, as also observed in $H_3S/D_3S$. The reason is analogous to that discussed for the lattice constant: owing to its larger mass, D atom is less affected by quantum anharmonic effects than H atom.

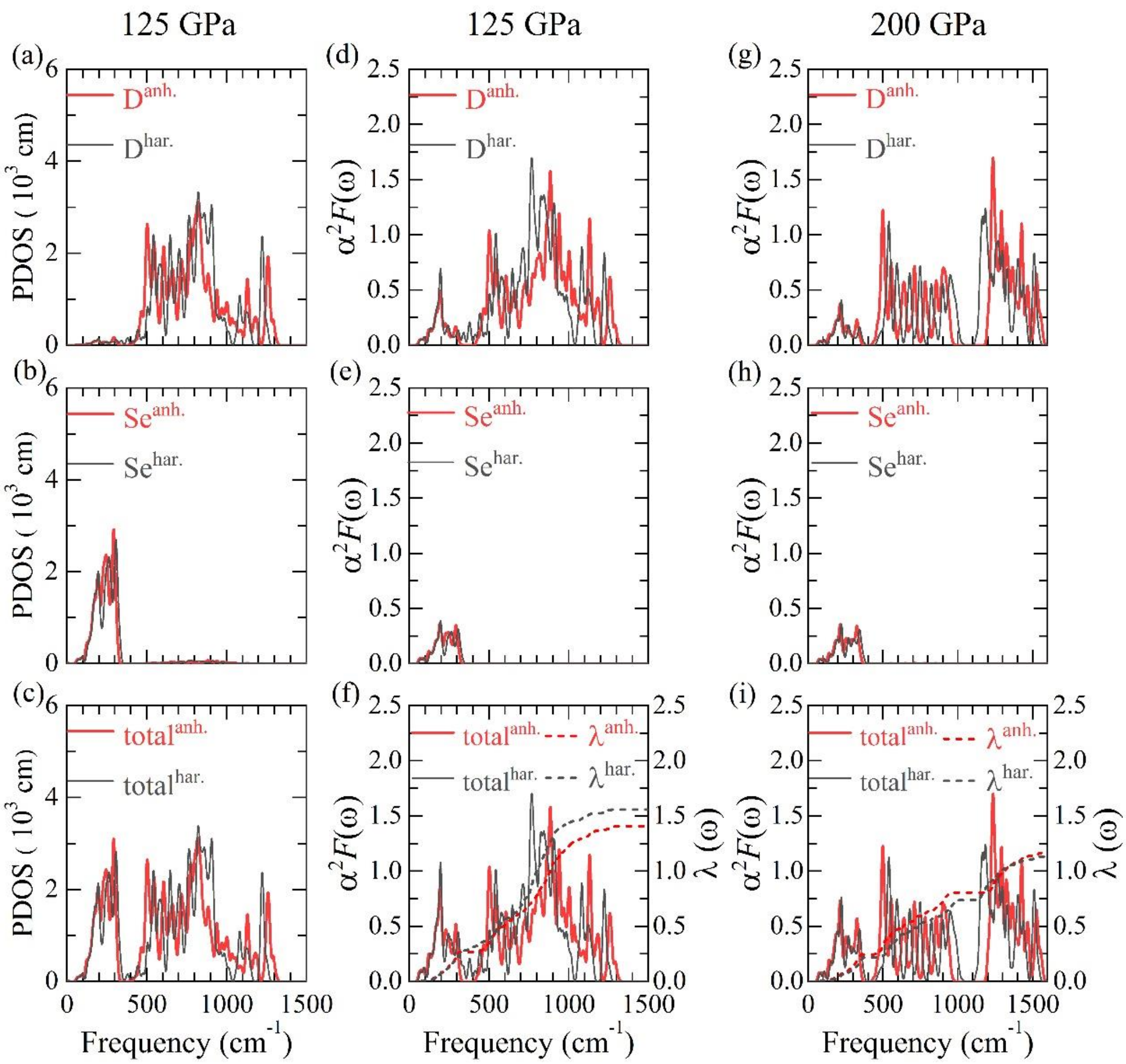

Fig. 4 (a-c) Projected phonon density of states (PDOS) of $D_3Se$ at 125 GPa: (a) D-atom projection, (b) Se-atom projection, and (c) total PDOS. (d-f) Projected Eliashberg spectral function $\alpha^2F(\omega)$ at 125 GPa: (d) D-atom projection, (e) Se-atom projection, and (f) total $\alpha^2F(\omega)$ together with the integrated electron-phonon coupling $\lambda(\omega)$. (g–i) Same as (d-f) but at 200 GPa. Black lines represent harmonic results, and red lines represent SSCHA (anharmonic) results. In panels (f) and (i), the dotted‑red and dotted‑black lines denote the anharmonic and harmonic integrated electron‑phonon coupling $\lambda(\omega)$, respectively.

For further exploration, the projected phonon density of states (PDOS) and projected Eliashberg spectral function $\alpha^2F(\omega)$ and its integral $\lambda(\omega)$, for the harmonic and anharmonic cases are shown in Fig. 4.

From Fig.4 (a-c), in the low-frequency region (below 400 cm$^{-1}$), the PDOS is dominated by the heavier Se atoms and is hardly affected by quantum anharmonic effects. This behavior differs from that of the Eliashberg spectral function $\alpha^2F(\omega)$. As shown in Fig. 4(d) and 4(g), the D atoms exhibit a contribution to $\alpha^2F(\omega)$ in the low-frequency region (below 400 cm$^{-1}$) that is of the same order of magnitude as that of Se, which is similar to our observations in $H_3Se$ [39] and $SnH_4$ [55].

Furthermore, comparing the harmonic and anharmonic (SSCHA) results, the integrated electron-phonon coupling $\lambda(\omega)$ plotted in Fig. 4(i) (for 200 GPa) clearly reveals the effects described earlier: the six mid-frequency bond-bending modes (400-900 cm$^{-1}$) are softened (except at the Γ point), while the three high-frequency bond-stretching modes (above 900 cm$^{-1}$) are hardened. Consequently, in the anharmonic calculation, $\lambda(\omega)$ gains a clear advantage in the middle-frequency range but gradually loses it at high frequencies. As a result, the final total $\lambda$ values from the harmonic and anharmonic calculations are similar, or even slightly larger in the harmonic case. In contrast, at lower pressure (e.g., 125 GPa, see Fig. 2(c) and 4(f)), the anharmonic $\lambda(\omega)$ shows an advantage in the middle-frequency range, while the anharmonic hardening of the high-frequency modes persists. As a consequence, the harmonic approximation yields a larger total electron-phonon coupling constant (1.56 for harmonic vs. 1.41 for anharmonic at 125 GPa). This is consistent with our findings for $H_3Se$ [39].

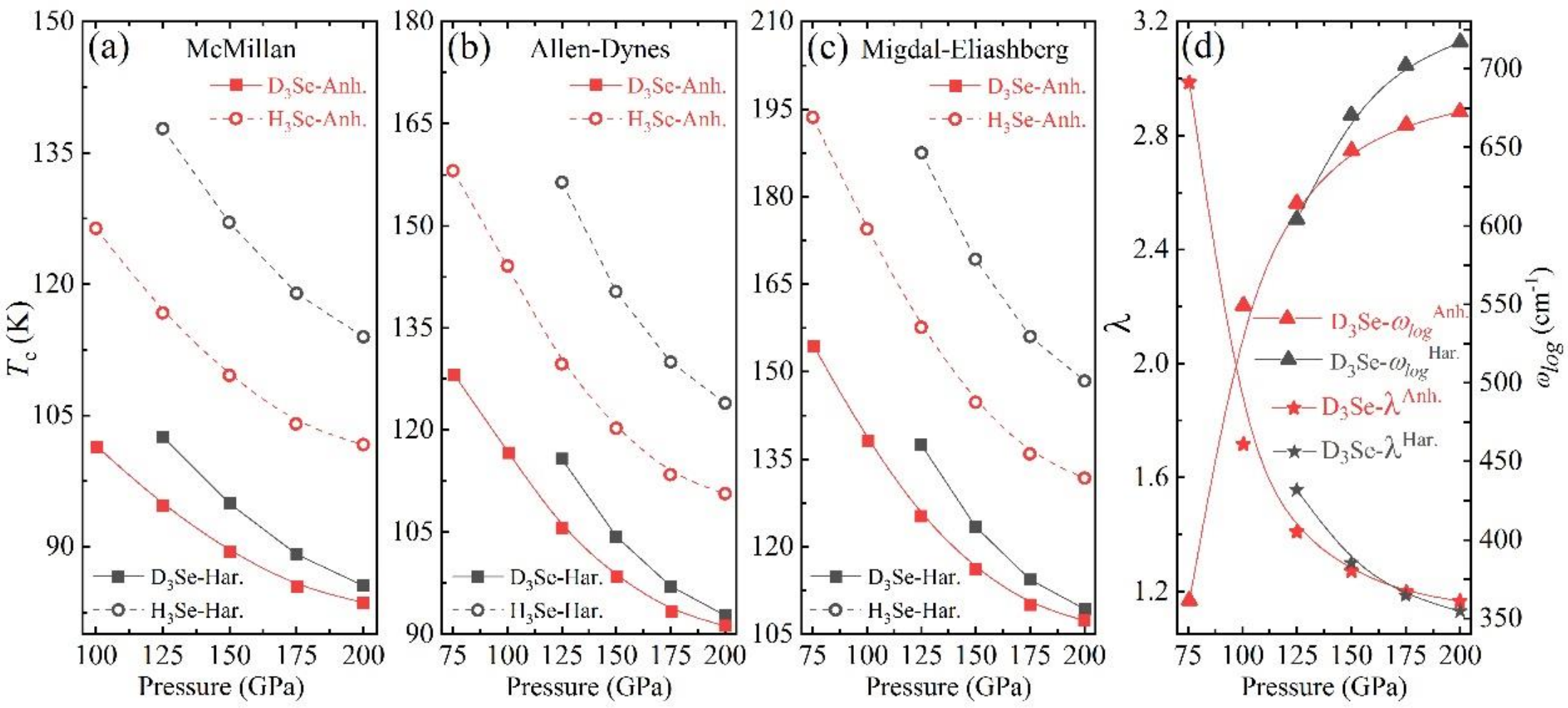

FIG. 5. (a-c) Superconducting critical temperature $T_c$ as a function of pressure for $H_3Se$ (circles) and $D_3Se$ (squares), calculated using (a) the McMillan equation, (b) the Allen-Dynes modified equation, and (c) the isotropic Migdal-Eliashberg equations. All calculations adopt a Coulomb pseudopotential $\mu*$=0.1. (d) Electron-phonon coupling constant $\lambda$ (left axis) and logarithmic average frequency $\omega_{\log}$ (right axis) for $D_3Se$ as functions of pressure. In each panel, black symbols and lines denote the harmonic (classical) results, while red symbols and lines correspond to the anharmonic (SSCHA) results. The upward triangle denotes $\lambda$, and the asterisk denotes $\omega_{\log}$.

The pronounced anharmonic renormalization of the optical phonons substantially influences the calculated superconducting critical temperatures. Fig. 5(a)-5(c) compare the superconducting critical temperature ($T_c$) obtained using (a) the McMillan equation, (b) the Allen–Dynes modified equation, and (c) the isotropic Migdal-Eliashberg equations, within both the harmonic approximation and the SSCHA framework (which incorporates anharmonic effects). For brevity, these are referred to as the harmonic $T_c$ and the anharmonic $T_c$, respectively. All calculations are performed with a Coulomb pseudopotential parameter $\mu*$=0.1. It should be noted that the anharmonic $T_c$ at 75 GPa is not included in Fig. 5(a), because the electron-phonon coupling constant $\lambda$ at this pressure is significantly larger than unity, rendering the McMillan equation inapplicable. As can be seen from the figures, the three methods yield qualitatively consistent results.

The $T_c$ for both $D_3Se$ and $H_3Se$ exhibits a monotonic decrease with increasing pressure in both harmonic and anharmonic calculations. For example, as shown in Fig. 5(c), the Migdal–Eliashberg equation predicts that the $T_c$ of $D_3Se$ decreases monotonically from 154 K at 75 GPa to 107 K at 200 GPa. This pressure-induced suppression of $T_c$ stems primarily from the general hardening of optical phonon modes under compression, and is consistent with the expected behavior for electron-phonon induced superconductivity.

Similar to $H_3Se$, the harmonic approximation slightly overestimates the $T_c$ for $D_3Se$ compared to the quantum anharmonic results. This behavior is consistent across systems with a highly symmetric bonding environment[36]. Interestingly, as shown in Fig. 5(d), the calculated $\lambda$ is not strongly suppressed by anharmonicity at high pressures (150~200 GPa). In fact, the anharmonic $\lambda$ at 200 GPa is even larger than its harmonic counterpart, a behavior that differs from systems like $AlH_3$[28]. This can be interpreted as a cancellation effect between the anharmonic softening of bond-bending optical modes and the hardening of bond-stretching modes. Despite the slight increase in $\lambda$ when anharmonicity is included, $T_c$ is still suppressed in the 175~200 GPa range. This suppression is attributed to other frequency renormalization effects captured within the anharmonic framework: $\omega_{log}$ is reduced from 702 to 665 cm$^{-1}$ at 175 GPa and from 717 to 673 cm$^{-1}$ at 200 GPa by the quantum and anharmonic effects.

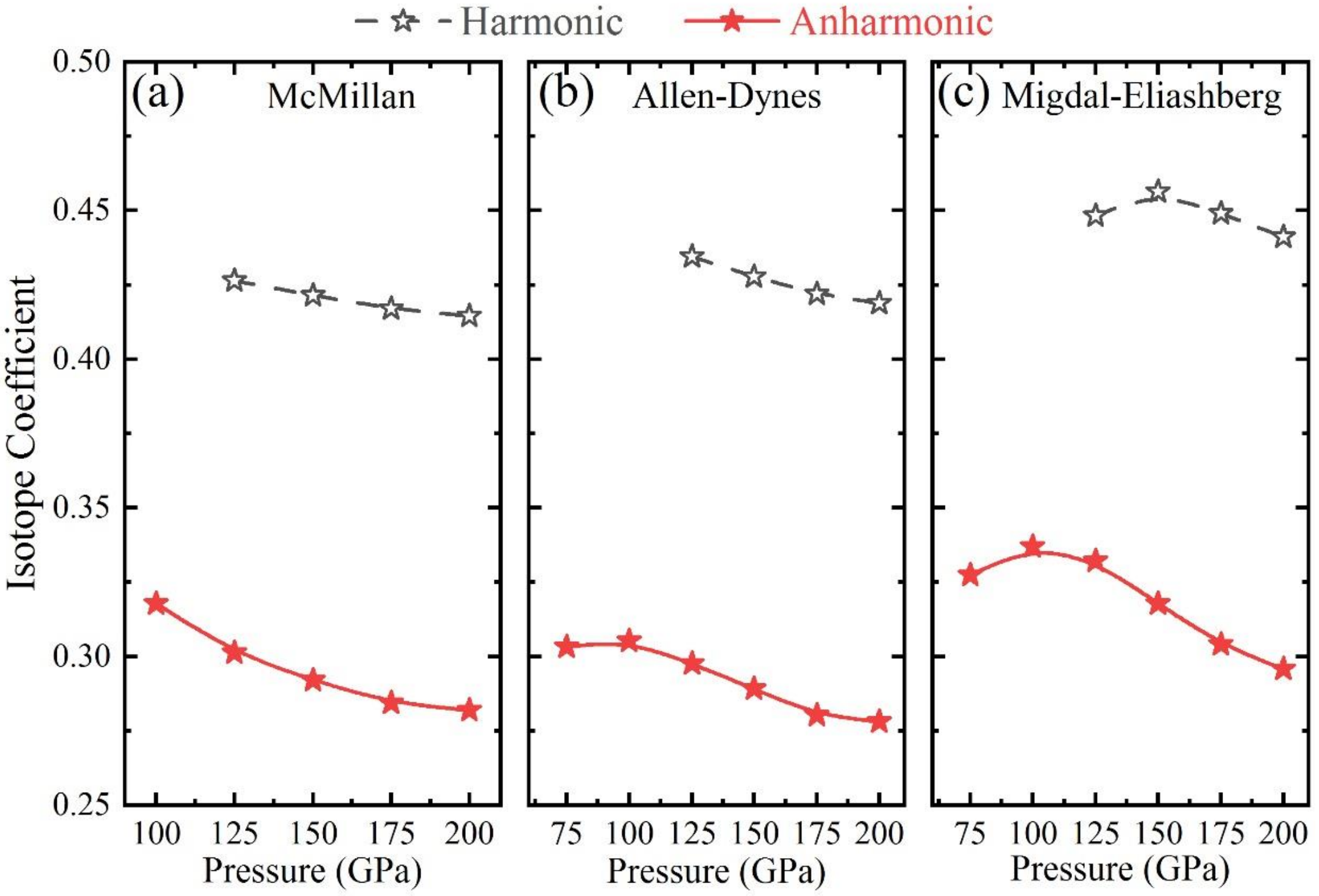

FIG. 6. Pressure dependence of the isotope coefficients $\alpha$ for $D_3Se$ and $H_3Se$. Red filled stars with solid lines and black open stars with dotted lines represent anharmonic and harmonic results, respectively. Calculations performed with (a) the McMillan equation; (b) the Allen-Dynes modified McMillan equation and (c) the Migdal-Eliashberg equation.

Fig. 6 (a-c) presents the calculated pressure-dependent isotope coefficient $\alpha$ for $M_3Se$ (M = H, D), derived from McMillan, Allen-Dynes and Migdal-Eliashberg equations, respectively. A key and consistent result across all equations is that the anharmonic $\alpha$ (red lines) remains below 0.35 in the studied pressure range (75 ~ 200 GPa), which is significantly lower than the canonical BCS value of 0.5 expected for a purely phonon-mediated single-element superconductor. A comparative analysis of the isotope coefficients calculated at the harmonic and anharmonic levels reveals a significant pressure-dependent divergence. As pressure increases, the disparity between the two theoretical approaches becomes increasingly pronounced, culminating at 200 GPa, where the harmonic approximation (e.g., $\alpha$=0.44 given by the Migdal-Eliashberg equations) yields an isotope coefficient that is one and a half times of its anharmonic counterpart ($\alpha$=0.29). This quantitative deviation underscores the necessity of incorporating anharmonic contributions to accurately describe the isotope effect under extreme compression conditions, as the harmonic picture progressively overestimates the isotopic response at higher pressures.

This finding aligns closely with the theoretical work by Errea et al. on $H_3S$ [32]. In that study, anharmonic calculations for $M_3S$ (M = H, D) yielded $\alpha \approx 0.35$ at 200 GPa, starkly contrasting with a much larger experimental estimate ($\alpha \approx 1.07$). Our results for the selenide analogue reinforce the theoretical prediction that anharmonic effects substantially suppress the isotope coefficient in high-pressure hydride/deuteride superconductors. The harmonic results (dashed lines in Fig. 6), which are closer to 0.5, likely overestimate $\alpha$ because they fail to capture the strong renormalization of phonon frequencies and electron-phonon matrix elements by quantum ionic motion.

Therefore, our calculated $\alpha \leq 0.35$ does not represent an "anomalous" suppression (as initially suggested for $H_3S$ based on experiment), but rather a theoretical suppression below the BCS value due to anharmonicity. Whether this is termed "anomalous" depends on the baseline expectation.

Compared to the simple BCS picture ($\alpha$=0.5), it is atypical. However, it appears to be a common feature predicted for strongly anharmonic, multi-component hydrides. The significant discrepancy between the anharmonic theory ($\alpha \leq 0.35$) and the large experimental $\alpha$ reported for $H_3S$ remains an open puzzle, possibly pointing to different stabilized structures for hydrides and deuterides in experiments. Our results for $D_3Se/H_3Se$ may provide a clear prediction for future experimental verification in selenide systems.

Furthermore, two factors cause the harmonic approximation to overestimate the isotope coefficient $\alpha$ for $D(H)_3Se$. First, at a given pressure the harmonic $T_c$ of $H_3Se$ is consistently higher than that of $D_3Se$, regardless of whether anharmonic corrections are included. Second, while anharmonic SSCHA corrections lower the $T_c$ for both isotopes, the reduction is less pronounced for $D_3Se$ because the heavier mass of D atom makes it less susceptible to quantum anharmonic effects. Consequently, after including anharmonic corrections, the $T_c$ of $H_3Se$ remains higher than that of $D_3Se$, but the difference between them reduces, leading to a smaller isotope coefficient compared to the harmonic prediction. By extension, if a hydride exhibited the opposite behavior—namely, an anharmonic increase in $T_c$ -the isotope coefficient could become larger after anharmonic corrections.

## Conclusion

In this work, we have systematically investigated the structural, vibrational, and superconducting properties of $D_3Se$ in the cubic $Im\bar{3}m$ phase under pressures from 60 to 200 GPa by combining first-principles calculations with the SSCHA. Our results demonstrate that quantum anharmonic effects play a decisive role, stabilizing the lattice down to ~66 GPa—well below the harmonic threshold—and inducing an effective internal pressure increase of ~6 GPa. The superconducting critical temperature $T_c$ reaches ~154 K at 75 GPa ($\mu^* = 0.1$, $\lambda = 3.0$) given by the Migdal-Eliashberg equations, confirming $D_3Se$ as a high-$T_c$ superconductor at moderate pressures. Most strikingly, anharmonicity dramatically reduces the isotope coefficient $\alpha$ to ~0.23 at 200 GPa, approximately half of the harmonic value. This result aligns closely with theoretical findings for the $H_3S$ system and reinforces the notion that anharmonicity fundamentally governs the isotope effect in hydrogen-rich superconductors. These findings underscore the indispensable role of anharmonicity and call for targeted experimental efforts in selenides to resolve the theory–experiment gap for the isotope coefficient, potentially guiding the search for ambient-pressure superconductivity.